\documentclass[12pt]{article}
\usepackage{setspace}
\def\pt{$p_T$}

\def\tt{$\cal T$}
\def\gev{GeV/$c$}
\def\bb{$\beta$}
\def\cc{$C(\beta,\eta)$}
\def\et{$\eta$}

\begin{document}

\begin{flushright}{OITS 759.7}\\[-0.8em]{NUC-MINN-04/9-T} \\
December 2004
\end{flushright}

\vspace*{.75cm}

\begin{center}
{\Large {\bf Forward Production in d+Au Collisions \\by Parton Recombination}}
\vskip .75cm
   {\bf   Rudolph C. Hwa$^1$,  C.\ B.\ Yang$^{1,2}$ and R.\ J.\ Fries$^3$}
\vskip.5cm

   {$^1$Institute of Theoretical Science and Department of
Physics\\ University of Oregon, Eugene, OR 97403-5203, USA\\
\bigskip
$^2$Institute of Particle Physics, Hua-Zhong Normal University,
Wuhan 430079, P.\ R.\ China
\bigskip
$^3$School of Physics and Astronomy, University of Minnesota, Minneapolis,
MN 55455}, USA
\end{center}

\begin{abstract}
Hadron production at forward rapidities in d+Au
collisions is studied in the framework of parton recombination in the
final state. Multiple scattering and gluon saturation in the initial
state are not explicitly considered. The recombination of soft and
shower partons is found to be important. The soft parton densities
are determined by extrapolation from the parametrization fixed at
$\eta=0$ with no unconstrained adjustments. The suppression of
$R_{CP}$ observed at high \et\ is understood as the simple
consequence of the reduction of the soft parton density on the
deuteron side compared to that on the gold side. The effect of
momentum degradation responsible for baryon stopping is also
considered. The asymmetry of backward-to-forward spectra can be
reproduced within the same framework without any change in the basic
physics.
\end{abstract}

\section{Introduction}

Particle production at forward rapidities in d+Au collisions has recently been
recognized as the fertile ground for testing models of hadron production that
embody diverse physical mechanisms.  Since no hot and dense medium is
created in d+Au collisions, they have generally been regarded as the type of
collision processes where the effects of final-state interaction (FSI) are
minimal, thus allowing the physics of initial-state interaction (ISI)
to manifest
itself most transparently \cite{lm}-\cite{dk}.  That is especially true at
forward rapidities where the saturation effect is expected to be important and
the physics of color glass condensate (CGC) may dominate
\cite{lm}-\cite{rb}.  Alternative approaches to the problem based primarily on
perturbative QCD (pQCD) have also been considered with emphasis on the
effects of nuclear shadowing \cite{rv,vg}.  The results do not provide
unambiguous interpretation of the data from BRAHMS \cite{ia}.  We consider
in this paper yet another approach that emphasizes the hadronization part of the FSI, 
and takes into account hard scattering in pQCD; however, instead of
fragmentation we consider the recombination of soft and shower partons.
Our treatment is an extension to the forward region of what has been found to
be successful at midrapidity for Au+Au collisions \cite{hy} and for d+Au
collisions \cite{hy2}, and is a natural mechanism for hadronization at
intermediate transverese momentum $p_T$.  It does not, in principle,
contradict the physics of CGC.  The issues are whether hadronization by recombination is important and  which one of the competing effects is
dominant in the kinematical region under examination.  Our results
indicate that the BRAHMS data \cite{ia} can be well reproduced in our
approach even when no adjustable parameters are used.

Experiments at the Relativistic Heavy Ion Collider (RHIC) have found that the
ratio, $R_{CP}$, of d+Au central to peripheral inclusive spectra for $1 < p_T
< 3$ \gev\ decreases monotonically from a value $\sim 1.8$ at
pseudorapidity $\eta \sim -2$ to a value $\sim 0.5$ at $\eta \sim 3.2$
\cite{ia,ml}.  This has led to the interpretation of a change of the
physics responsible for the phenomena from the gold side ($\eta <0$) to the
deuteron side ($\eta >0$) \cite{lm}.  For $\eta \leq 0 $ the enhancement of the
particle yield at intermediate $p_T$ with respect to binary collision scaling,
i.e., the Cronin effect \cite{jc}, is generally regarded as the
result of multiple
scattering in ISI \cite{xw,aa}.  For $\eta >0$ saturation physics is considered
to be dominant, especially at large $\eta$, so that there is suppression,
instead of enhancement, in particle production \cite{dk,dkk,rb}.  Neither
effect takes into account any details about hadronization in FSI.  Such explanations of the data
cannot account for the phenomenological fact that $R^p_{CP}$ (for protons) is
larger than $R^{\pi}_{CP}$ (for pions) at intermediate $p_T$ \cite{fm}.
Hadronization of partons to a proton or a pion is an issue that involves the
final state, and the use of fragmentation functions (FF) appropriate for the
produced hadrons in the usual factorizable way would necessarily yield a
$p/\pi$ ratio $<1$ by virtue of the nature of the FFs, contrary to what has
been observed \cite{sa}.

The subject of large $p/\pi$ ratio has been addressed by several groups that
consider parton recombination as the hadronization process in the final state
in Au+Au collisions \cite{hy3,gkl,fmnb}.  For hadrons produced at
intermediate $p_T$ it is the recombination of thermal partons at lower
$p_T$ that gives rise to the higher yield of the baryons compared to mesons.
In Ref. \cite{hy} the additional component of the recombination of thermal
and shower parton is introduced, and is shown to be dominant in the
$3<p_T<9$ \gev\ range.  That component turns out to be crucial to explain
the Cronin effect, when the formalism is extended to d+Au collisions, without
the need for any $k_T$ broadening in ISI \cite{hy2}.  Although there are no
thermal partons in d+Au collisions as in Au+Au collisions, there are soft
partons that play the same role.  It is found that the property $R^p_{CP} >
R^{\pi}_{CP}$ at intermediate $p_T$ can readily be reproduced in the
recombination model \cite{hy2,hy4}.  Based on the success of that
description of particle production at midrapidity in d+Au collisions, we now
extend the treatment to the forward region $(\eta >0)$ without the
introduction of any new physics.  This
extension should be considered whether or not the signature of new physics is
present in the forward region, since the contribution from conventional
physics forms the background that must be understood to facilitate the
identification of any new signal.  As we shall show, it seems that
recombination in FSI is completely adequate to explain the suppression at
large $\eta$.

The main reason for $R_{CP}$ to decrease with increasing $\eta$ is simple.
Since the recombination process that involves soft partons is important, the
dependencies of the soft parton density on rapidity and centrality directly
affect the particle spectra at intermediate $p_T$.  That density that is
reflected in $dN/d\eta$ is known to decrease with increasing $\eta$ with the
consequence that $R_{CP}$ also decreases with $\eta$.  Since CGC has
successfully reproduced the hadron multiplicity $dN/d\eta$, one may take
the view that the suppression at large $\eta$ has its origin in saturation
physics.  That may be the case. From the point of view of FSI the suppression is a direct result of parton recombination involving soft partons whose density decreases with $\eta$; indeed, our treatment of hadron production is compatible with any model that can generate the soft parton spectra that we shall determine phenomenologically.

Before we consider hadron production quantitatively, let us make some general remarks here on recombination. Since the formation of a pion by a $q\bar q$ pair implies the loss of degrees of freedom, it is sometimes thought that the recombination process leads to a decrease of entropy. However, the entropy principle should  not be applied locally.  A global consideration must recognize that the bulk volume is increasing during the hadronization process, and thus compensates the decrease of entropy density that is a local quantity. Furthermore, in the $q\bar q\to\pi$ process color singlet is formed by color mutation of the quarks through soft gluon radiation that carries the extra degrees of freedom without significantly altering the local relation that the sum of the parton momenta equals the pion momentum. All gluons hadronize by conversion to $q\bar q$ pairs first, so there is a cascading of the degrees of freedom to lower and lower momentum region where the pion multiplicity accumulates.  Eventually all degrees of freedom are converted from partons to hadrons. Such non-perturbative processes cannot be calculated, and our formalism does not contain explicitly the feed-down process of partons cascading to the very low $p_T$ region before recombination. Indeed, such a process need not be made explicit in our approach, since we do not determine the soft parton distribution in a model of evolution from the initial state. Instead, 
we determine the soft parton distribution from the observed pion distribution in the $0.5<p_T<2$ GeV/c region. The soft partons thus obtained are defined in the context of recombination. We use them in the same context when we consider their recombination with shower partons. Thus our procedure of treating hadronization at all $p_T>0.5$ GeV/c is totally self-consistent.

It is also appropriate at this point to make another general remark about our treatment of recombination. In a one-dimensional (1D) formulation of the recombination process in momentum space, one may question whether we are ignoring the spatial extent of the recombining subsystem normal to the collinear momentum vectors of the quarks and hadrons. Partons that are separated by a distance of the order of the transverse size of the bulk volume but have parallel momenta are not likely to recombine. That is indeed an issue that has to be faced in a model in which the spatial and momentum distributions of the partons are generated from the initial state and evolve according to some sensible dynamics. However, that is not what we do, as we have already stated at the end of the preceding paragraph.  For the soft parton distribution, which is the crux of the issue here, we start from the observed soft pion distribution.  For every $\vec p$ of such a pion we claim that it can only arise from the recombination of a quark at $\vec p_{1}$ and an antiquark at $\vec p_{2}$ that are not only collinear in momentum vectors, but also spatially overlapping within a transverse space of the order of a hadron. In other words, since we determine the soft parton distribution from the final state, the result automatically implies that only those partons can recombine to give the observed hadron. The 1D formulation of recombination is therefore appropriate for the way in which it is applied.

\section{Pion Production at $\eta = 0$}

We summarize first the production of pions by parton recombination at
midrapidity in d+Au collisions and make minor adjustments to align our
calculation for the centrality cuts of the BRAHMS experiment \cite{ia}.  We use
the formalism for hadronization described in Refs. \cite{hy,hy2}, where the
inclusive distribution of pion at $\eta = 0$ in a 1D description is given by
\begin{eqnarray}
p{dN_{\pi}  \over  dp} = \int {dp_1 \over p_1}{dp_2 \over p_2}F_{q\bar{q}'}
(p_1, p_2) R_{\pi}(p_1, p_2, p),
\label{1}
\end{eqnarray}
$p$ being in the direction of the detected pion.  $F_{q\bar{q}'}(p_1, p_2)$
is the joint distribution of a $q$ and a $\bar{q}'$ at $p_1$ and $p_2$, and
$R_{\pi}(p_1, p_2, p)$ is the recombination function for forming a pion at
$p$:  $R_{\pi}(p_1, p_2, p) = (p_1 p_2/p) \delta(p_1+ p_2 -p)$.  For $p$ in
the transverse plane so that $p_T = p$, the distribution
$dN_{\pi}/d^2p d\eta|_{\eta =0}$, averaged over all $\phi$, is
\begin{eqnarray}
{dN_{\pi}  \over  pdp} =  {1 \over p^3} \int^p_0 dp_1F_{q\bar{q}'} (p_1,
p-p_1) .
\label{2}
\end{eqnarray}
$F_{q\bar{q}}$ has three components
\begin{eqnarray}
F_{q\bar{q}'} = {\cal TT} + {\cal TS} + {\cal SS}\ ,
\label{3}
\end{eqnarray}
where ${\cal T}$ stands for soft parton distribution and ${\cal S}$ for shower
parton distribution.  For Au+Au collisions ${\cal T}$ would refer to the
thermal partons \cite{hy}, but in d+Au collisions the notion of thermalization
is inappropriate, but soft partons nevertheless exist.  Since they are treated
in the same way, the same notation is used with ${\cal T}$, which can
be regarded as
   referring to the last
letter of ``soft.''  At low $p_T$ the observed pion distribution is
exponential,
which suggests the form
\begin{eqnarray}
{\cal T}(p_1) = p_1{dN^{{\cal T}}_q  \over  dp_1}= Cp_1 \exp (-p_1/T),
\label{4}
\end{eqnarray}
so that the ${\cal TT}$ component in Eq.\  (\ref{3}) yields
\begin{eqnarray}
{dN^{{\cal TT}}_{\pi}  \over  pdp} =  {C^2 \over 6} \exp (-p/T) \ .
\label{5}
\end{eqnarray}
The values of $C$ and $T$ have been determined in Ref.\ \cite{hy2}
already from the low-\pt\ d+Au
data. Since our approach to hadron production at intermediate \pt\ at
RHIC is to emphasize parton
recombination in the final state, we use the phenomenological input
for the soft component,
  without relying on any specific model for
soft partons so that our result will be independent of the
reliability of such models.

The distribution $\cal S$\ is a convolution of the hard parton
distribution $f_i(k)$ with transverse
momentum $k$ and the shower parton distribution (SPD) $S_i^j(z)$ from
hard parton $i$ to semi-hard
parton $j$
\begin{eqnarray}
{\cal S}_j(p_1)&=&\sum_i\int_{k_{\rm min}}dk\,kf_i(k)\,S_i^j(p_1/k)\ ,
\label{6}
\end{eqnarray}
where $k_{\rm min}$ is set at 3 \gev, below which the pQCD derivation
of $f_i(k)$ is invalid. For each $i
$, $f_i(k)$ depends on the parton distribution functions, nuclear
shadowing, and hard scattering
cross sections. The result is presented in the power-law form, whose
parameters are tabulated in
Ref.\ \cite{hy2}. The SPDs are obtained from the FFs and are given in
Ref.\ \cite{hy5}.

It is sufficient with the above specification of \tt\ and $\cal S$\ to
calculate the \pt\ spectrum of pion
by use of Eq.\ (\ref{2}). However, for comparison with the BRAHMS
data we need the values of $C$
for the corresponding centrality cuts. Let us use \bb\ to denote the
centrality cut, which is an
experimental quantity related to the impact parameter $b$. In Ref.\
\cite{hy2} it is found that
$C(\beta,\eta)$ varies with \bb\ at $\eta=0$ according as
$C(\beta,0)=12, 11, 7.8, 5.65$ (GeV/$c)^{-1}$ for
$\beta=$ 0-20, 20-40, 40-60, 60-90\%, respectively. The centrality
cuts of BRAHMS \cite{ia} are
\bb=0-20, 30-50, and 60-80\%. We therefore make the interpolation and
set $C(\beta,0)=12, 9.0,$ and
6.55 (GeV/$c)^{-1}$, respectively. The value of the inverse slope is
$T=0.208$ \gev, as determined in Ref.\
\cite{hy2}.

Before showing the result, we remark that the $\cal S$$\cal S$\ component in Eq.\
(\ref{3}) corresponds to fragmentation, if the two shower partons
originate from the same hard
parton, i.e.,
\begin{eqnarray}
({\cal S}_j{\cal
S}_{j'})(p_1,p_2)&=&\sum_i\int_{k_{\rm min}}dkkf_i(k)
\left\{S_i^j\left(p_1\over k\right),S_i^{j'}\left({p_2\over
k-p_1}\right)\right\}\ ,
\label{7}
\end{eqnarray}
where the curly brackets signify the symmetrization of the leading
parton momentum fraction
\cite{hy, hy5}. If the two shower partons are from two independent
jets, then there would be two
$f_i(k)$ distributions. The recombination of such partons is very
unlikely to occur in d+Au
collisions at 200 \gev\ and will be ignored.

The results of our calculation for $\pi^+$ production are shown in
Fig.\ 1 for the three
centralities, displaced by factors of $10^2$ from neighboring ones.
The light solid lines show the
soft-soft (\tt\tt) components that are straight lines in the log plot.
The dashed lines show the soft-shower (\tt$\cal S$) contributions and the
dash-dotted lines the
shower-shower ($\cal S$$\cal S$) contributions.  The heavy solid lines are the
sums, whose deviations from the
straight lines are indicative of the effects of hard scattering. Note
that the \tt$\cal S$\ contribution
becomes less important as \bb\ increases because $C(\beta,0)$
decreases. Indeed, if the \bb=60-80\%
case is regarded as being almost like the $pp$ collision, we see from
Fig.\ 1 that the neglect of
the \tt$\cal S$\ contribution does not constitute a bad approximation, and
the large \pt\ behavior is
essentially governed by jet fragmentation, as has traditionally been
used to treat $pp$ collisions.
However, that is not the case for central d+Au collisions. The
\tt$\cal S$\ contribution to the spectra
is what accounts for the Cronin enhancement at intermediate \pt\
without $k_T$ broadening due to ISI
\cite{hy2}. The \pt\ distributions have been shown to agree with the
PHENIX data \cite{fm}. The
$R_{CP}$ ratio for the \bb\ values of the BRAHMS data will be shown
in Fig.\ 5 below.

\section{Pion Production at $\eta>0$}

Since the formalism for pion production at $\eta=0$ described in the
preceding section successfully
reproduces the  experimental \pt\ spectra at all centralities, we now
make a straightforward
extension to the $\eta>0$ region. This extension should be made with
no change in the basic physics
underlying the  formalism in order to provide a baseline for
comparison with the data before the
search for the signature of any other physical origin.

The  quantities that must be modified for the $\eta>0$ region are
$C(\beta, \eta)$ for \tt\ and
$f_i(k)$ for $\cal S$.  For the inverse slope $T$ we shall proceed in two
steps. First, we keep $T$ fixed as \et\ is increased, since no data
on low-\pt\ $\pi^+$ spectra are available to serve as our guide for
its modification.  It is of interest to see how close the calculated
$R_{CP}$ will turn out to be in comparison to the data, when the
constant-$T$ assumption is applied for the purpose of introducing no
adjustable parameter. Later, we shall allow $T$ to depend weakly on
\bb\ and \et\ and show that the fit of $R_{CP}$ can be improved.

Since the observed rapidity
density $dN_{\rm ch}/d\eta$ is an integral over the \pt\
distribution, which is dominated by the soft contribution at low \pt,
we see from Eq.\ (\ref{5}) that $dN_{\rm ch}/d\eta$
   should be
proportional to $C^2(\beta, \eta)$. We can therefore determine
$C(\beta,\eta)$ by use of the formula
\begin{eqnarray}
C(\beta,\eta)=C(\beta,0)\,\left[{dN_{\rm ch}/d\eta(\beta)\over dN_{\rm
ch}/d\eta|_{\eta=0}(\beta)}\right]^{1/2}\ .
\label{8}
\end{eqnarray}
$C(\beta,0)$ is given in the preceding section, while $dN_{\rm
ch}/d\eta(\beta)$ is known from
PHOBOS data \cite{rn}. We thus obtain the values of \cc\ as shown in
Table I. Since PHOBOS does not
have exactly the same centrality cuts as BRAHMS, some interpolation
between neighboring values
have been made to deduce the numbers in Table I.

\begin{table}
\caption{Values of $C(\beta,\eta)$ in (\gev)$^{-1}$}
\begin{center}
\begin{tabular}{|r|c||cccc|} \hline

  && &\quad\quad  $\eta$&&\\ \hline
   &&$0$&$1$&$2.2$&$3.2$\\ \hline\hline
&0-20\%&12.0&11.1&9.01&7.05\\
$\beta$ &30-50\%&9.0&8.5&7.9&6.0\\
&60-80\%&6.55&6.6&6.1&5.1\\ \hline
\end{tabular}
\end{center}
\end{table}

We have parametrized the hard parton distributions $f_i(k_T,y)$,
as before \cite{hy2,ds}, in the power-law form
\begin{eqnarray}
f_i(k_T,y)\equiv {1\over \sigma_{\rm in}}\,{d\sigma_i^{d+Au}\over
d^2k_T\,dy}=K\,A_i\left(1+{k_T\over B_i}\right)^{-n_i}
\label{9}
\end{eqnarray}
for $y\le 1$. For larger values of $y$ the spectra are increasingly 
suppressed at high $k_T$ because of the phase space boundary that 
requires $k_T < k_0(y)$.
The kinematic limit is given by $k_0(\eta) = \sqrt{s}/(2\cosh y)$
and takes the values 21.9
and 8.13 GeV/$c$ for $y=2.2$ and 3.2 respectively. To take into account the
change of $f_i(k_T,y)$ from positive to negative curvature around
$k_T \approx 0.5 \, k_0$,
equation (\ref{9}) must be modified by a dampening factor. Therefore we use
the parametrization
\begin{eqnarray}
f_i(k_T,y)=K\,A_i\left(\pm 1+{k_T\over
B_i}\right)^{-n_i}\,\left(1-{k_T\over k_0}\right)^{m_i}\ ,
\label{10}
\end{eqnarray}
for $y>1$
where the $\pm1$ sign is used in accordance to whether the given
value of $B_i$ is preceded by a +
or $-$ sign.

The parametrizations are obtained from leading order
minijet calculations using CTEQ5 parton distributions \cite{Lai:1999wy}.
EKS98 shadowing \cite{Eskola:1998df} was used for the Au nucleus while the
deuteron was treated as a superposition of a proton and a neutron
without further nuclear modifications.
The values of all the parameters for central d+Au
collisions (\bb\ = 0-20\%) are given in Table II, corresponding to
$\sigma_{\rm
in}=40.3$ mb and the $K$ factor unspecified. For other centralities
scaling in the number of binary collisions, $N_{\rm coll}$, is
assumed. We set $K=2$ in our calculation below.

\singlespacing

     \begin{table}
\caption{Parameters in Eqs.\ (\ref{9}) and  (\ref{10}) }
\begin{center}
\begin{tabular}{|l||c|c|c|c|c|c|c|c||} \hline
$\quad\quad y$&$i$ &$g$&$u$&$d$&$s$&$\bar{u}$&$\bar{d}$\\ \hline\hline
&A &196.52&55.65&60.74&3.114&11.55&12.23\\
\quad -0.75&B &1.442&1.064&1.045&1.657&1.330&1.292\\
&n&8.654&7.533&7.483&8.798&8.385&8.319\\ \hline
&A &254.06&61.64&65.26&3.953&13.35&13.27\\
\quad -0.25&B &1.265&0.996&0.990&1.456&1.228&1.223\\
&n&8.207&7.314&7.293&8.320&8.102&8.101\\ \hline
&A &244.63&59.51&61.57&3.786&13.02&12.50\\
\quad\ 0.25&B &1.260&0.991&0.993&1.453&1.218&1.229\\
&n&8.175&7.281&7.280&8.292&8.068&8.099\\ \hline
&A &177.86&51.35&51.93&2.745&10.88&10.34\\
\quad\ 0.75&B &1.419&1.039&1.050&1.646&1.291&1.308\\
&n&8.546&7.420&7.439&8.711&8.263&8.308\\ \hline
&A &132.78&43.64&43.70&2.030&9.016&8.580\\
\quad\ 1.0&B&1.600&1.103&1.120&1.874&1.384&1.402\\
&n&8.959&7.590&7.621&9.192&8.496&8.544\\ \hline
&A &12460&5.68e6&1.77e7&65.35&6.562e12&1.045e8\\
\quad\ 2.2&B&0.3184&(-)0.03396&(-)0.0277&0.4420&0.2662e-3&0.02054\\
$(k_0=21.9)$&n&5.939&4.873&4.897&5.900&5.191&5.241\\
&m&7.000&5.320&5.341&7.657&5.966&5.951\\ \hline
&A &10080&2.391e5&2.349e5&30.396&485.45&814.80\\
\quad\ 3.2&B&0.3360&(-)0.05117&(-)0.05218&0.6214&(-)0.1410&(-)0.1287\\
$(k_0=8.13)$&n&5.977&4.539&4.555&6.545&4.574&4.607\\
&m&6.024&4.548&4.559&6.189&4.943&4.948\\ \hline
\end{tabular}
\end{center}
\end{table}
\doublespacing

Using \cc\ in Eq.\ (\ref{4}) and $f_i(k_T,y)$ in Eq.\ (\ref{6}),
neglecting the difference between
$\eta$ and $y$ for $\eta\ge 1$, and fixing $T$ at 0.208 \gev\ as for
\et=0, we can now calculate
$dN_{\pi^+}/pdpd\eta$ according to Eqs.\ (\ref{2}) and (\ref{3}) and
obtain the results shown in
Fig.\ 2 for \bb\ = 0-20\% and 60-80\%, that for 30-50\% being in
between the two. Clearly, the \pt\
distributions are affected by the increase of $\eta$ mainly in the
large-\pt\ region. There are smaller changes at low \pt\ as seen in
the log scale, although they are not negligible in the linear scale,
since
the spectra there are proportional to $C^2(\beta,\eta)$ that varies
substantially with $\eta$
according to Table I. The $\eta$ dependence of the \pt\ distributions
in Fig.\ 2 is our prediction
for which we have not adjusted any free parameters. No data on
identified pions are currently available
to check those results.

The notable feature of Fig.\ 2 is that at $\eta=3.2$ the \pt\
distributions behave nearly as
straight lines in the log plot. The exponential behavior suggests
that only the soft partons
contribute to the pion formation. To see this more clearly, we show
in Fig.\ 3 the different
contributions to the spectra for \bb\ = 0-20\% and $\eta=3.2$.
Indeed, the \tt$\cal S$\ and $\cal S$$\cal S$\
components are much smaller than the \tt\tt\ component, and are
insignificant for $p_T<3$ \gev. The
reasons are twofold: not only is the soft parton density lower at
$\eta=3.2$, but also the hard
parton distributions are severely suppressed at high $k_T$. The
former is evident in Table I; the
latter is not as obvious in Table II. We plot the gluon distributions
in Fig.\ 4 for the four values
of $\eta$, and see the precipitous fall for $\eta=3.2$, as $k_T$
approaches the kinematical limit at
$k_T=8.13$ \gev. The suppression of $f_i(k_T,\eta)$ does not reduce
the $\cal S$\ term quadratically
because $f_i(k_T,\eta)$ appears only once in Eq.\ (\ref{7}).
Consequently, the \tt$\cal S$\ and $\cal S$$\cal S$\
components can have comparable magnitudes in Fig.\ 3. Their
significantly reduced contribution to
the overall distribution exposes the \tt\tt\ contribution to be the
dominant component for \pt\ up to 3
\gev. One can reasonably question the validity of extrapolating the
soft parton distribution
${\cal T}(p_1)$ to $p_1\sim 1.5$ \gev\ in its exponential form. Our
view is that, instead of
adopting some low-\pt\ model that has its own ambiguities, it is
sensible to use the exponential
form of Eq.\ (\ref{4}) for the soft parton distribution as a working
hypothesis without introducing
extra free parameters so as to make predictions that can be tested
experimentally.
The important observation is that the hard partons are suppressed at
high $\eta$ and that any
prediction by pQCD should not neglect the soft background, which is
shown to be more important than
fragmentation at high $\eta$.

Having obtained the pion spectra at all \bb\ and \et, we can now
calculate the central-to-peripheral
ratio
\begin{eqnarray}
R_{CP}(\beta,\eta)={dN_{\pi}/p_Tdp_Td\eta(\beta)/\left<N_{\rm
coll}(\beta)\right>\over
dN_{\pi}/p_Tdp_Td\eta(\beta_p)/\left<N_{\rm coll}(\beta_p)\right>}\ ,
\label{11}
\end{eqnarray}
where the reference $\beta_p=$ 60-80\% and $\left<N_{\rm
coll}(\beta)\right>$ is the average number of binary collisions at
\bb. The results for \bb\ = 0-20\% and 30-50\% are shown in Fig.\ 5
for the four
values of \et. The data points are for $(h^++h^-)/2$ for $\eta=0$ and
1, and for $h^-$ for $\eta=2.2$ and 3.2 \cite{ia}. The case of \et\ =
0 shows the Cronin effect
that is well described by our result where the solid line is for \bb\
= 0-20\% and the dashed line
for \bb\ = 30-50\%. As in Ref.\ \cite{hy2}, no $k_T$ broadening by
multiple scattering in ISI has
been put in. For $\eta\ge 1$, although the agreement of our results
with the data is not perfect,
they nevertheless exhibit the essence of the trend, i.e., $R_{CP}$
becomes smaller as \et\ is
increased. That feature has been regarded as the distinctive
characteristics of forward production,
and is now approximately reproduced by our treatment that contains no
new physics and no adjustable
parameters. The case of $\eta=3.2$ is the simplest to interpret,
since the shower contribution is
insignificant. The constancy of $R_{CP}$ in our result for $p_T<3$
\gev\ is a consequence of the fact that we have
fixed the value of the inverse slope $T$ for the soft parton
distribution, independent of
centrality. The suppression of $R_{CP}$ at $\eta=3.2$ is due to the
decrease of \cc\ with increasing
\et\ and the insufficiently fast decrease of \cc\ with increasing
\bb\ to overcome the decrease of
$\left<N_{\rm coll}(\beta)\right>$ that rescales the spectra in Eq.\
(\ref{11}). In short, since the
density of soft partons diminishes as one goes far into the deuteron
side, less particles are
produced by the recombination of those soft partons.

In the foregoing we have fixed $T$ for all $\beta$ and $\eta$ as an
assumption  for the sake of not making it an adjustable parameter.
The theoretical results, as shown in Fig.\ 5, are remarkably close to
the data. However, a mild dependence of $T$ on $\beta$ and $\eta$
cannot be excluded. Indeed, the increase of $R_{CP}$ with $p_T$ in
the data at $\eta=3.2$ suggests a decrease of $T$ with $\beta$. We
adopt a simple parametrization of that dependence as
follows
\begin{equation}
T(\beta,\eta) = T_0(1 -\varepsilon\beta\eta)
.  \label {11a}
\end{equation}
Since $R_{CP}$ is plotted in linear
scale in Fig.\ 5, it is possible to determine $\varepsilon$ by
fitting the data at $\eta=3.2$, despite the absence of the spectra
themselves  (which would be in log scale and insufficiently accurate
to determine small differences in $T$ by themselves separately). We
fix $T_0=0.208$ GeV (more precisely, $T_0^{-1}=4.8$/GeV), and set
$\beta=0.1, 0.4,$ and 0.7 for 0-20\%, 30-50\% and 60-80\% centrality,
respectively. We vary $\varepsilon$ to obtain the best fit of the
data in open circles at \et=3.2 in Fig.\ 5, and get

\begin{equation}
\varepsilon=0.0205 .
\label{12a}
\end{equation}
The results for all other values of \bb\
and \et\ are shown in Fig.\ 6. Compared to Fig.\ 5, there is a slight
improvement of the agreement with data at \et=1.0, but is a little
higher than the data at \et=2.2 for $p_T>2.5$ GeV/c.  Generally
speaking, the trend of the data with increasing \pt\ is better
reproduced when $T(\beta,\eta)$ is allowed to decrease slightly with
\bb\ and \et. That decrease is  less than 4.6\% only even at the
highest values of \bb\ (0.7) and \et\ (3.2). Thus the constant $T$
assumption is not a bad approximation and serves to reproduce the
data reasonably well as  in Fig.\ 5. However, we have not yet
exhausted all aspects of physics that can influence the fit of the
data.

\section{Momentum Degradation}

There is a piece of physics that we have not yet considered, but it
is a phenomenological fact that
should not be ignored. Baryon stopping generally refers to the loss
of projectile proton momentum in $pA$ collisions,  as
it passes through a target nucleus. Although such nomenclature is
misleading from the point of view
of the role that the proton constituents play, empirical evidence for
the momentum degradation of
the detected nucleon as a function of the nuclear size is not
disputed. A number of experiments have
shown that the produced nucleon distribution in $pA$ collisions has
the exponential form in $x_F$
\begin{eqnarray}
{dN_N\over dx_F} \propto {\rm exp} [-\Lambda(\nu)\,x_F]\ ,   \label{12}
\end{eqnarray}
where the slope $\Lambda(\nu)$ depends on the average number of
collisions $\nu$ \cite{bc,sva}.
Baryon stopping loosely refers to the phenomenon that $\Lambda(\nu)$
increases with $\nu$. Our
question here is whether such a behavior has a dynamical origin that
can affect our treatment of
forward production in d+Au collisions. The question is relevant,
since in both problems there is
suppression of production probability at high \et.

The production of leading nucleon in $p+A$ collisions has been
studied in the framework of the
valon model for low-\pt\ processes \cite{rch}, and the distribution
in Eq.\ (\ref{12}) is obtained
by attributing the momentum degradation effect to the projectile
valons as they traverse the target
nucleus \cite{hy6}. That is not contradictory to the information
gained from the more recent
experiments at RHIC, where energy loss effects are found to be absent
at large \pt\ in d+Au
collisions \cite{bbb, ssa, ja, ia2}. The former problem is at low
\pt\ and valons are dressed
valence quarks of the proton, whereas the latter refers to hard
partons that go through the nucleus
at large angles with negligible interaction with the cold medium. It
is the interpolation between
these two extremes that is pertinent to the \et\ dependence in our
problem here.

The nucleon distribution in Eq.\ (\ref{12}) has been converted in the
valon model to the pion
distribution in the form
\begin{eqnarray}
{dN_{\pi}\over dx_F} \propto {\rm exp} [-\lambda(\nu-1)\,x_F]\ ,   \label{13}
\end{eqnarray}
where $\lambda=0.2$ \cite{hy7}. There is no reliable way to relate
that behavior in $x_F$ at low
\pt\ to the \et\ dependence at intermediate \pt\ without treating the
transition from soft to hard
processes. Since the boundary condition is that there is no energy
loss at $\eta=0$, we adopt the
ansatz that the degradation factor is
\begin{eqnarray}
\zeta(\beta,\eta)={\rm exp} [-\kappa (N_c-1)\eta]\ ,   \label{14}
\end{eqnarray}
where $N_c=\left<N_{\rm coll}(\beta)\right>$ and $\kappa$ is a
parameter to be determined from
$\lambda$ by matching Eqs.\ (\ref{13}) and (\ref{14}) at forward
rapidity. The expression for
$\zeta(\beta,\eta)$ represents the property that the larger \et\ is,
the more time the constituents
of the projectile spend in the valon state, while the valons
propagate through the nuclear medium
and suffer momentum degradation.

To relate Eqs.\ (\ref{13}) and (\ref{14}) we note that at $\eta=3.2$,
if $\left<p_T\right>=2$ \gev,
the corresponding $\left<x_F\right>=0.25$. For $\beta=$ 0-20\%, we
use the values $\nu\approx 9$
(for $p$Au collisions) and $N_c\approx 15$ (for $d$Au collisions), and get
\begin{eqnarray}
\kappa\approx 0.01\ .   \label{15}
\end{eqnarray}
We now use this value of $\kappa$ in Eq.\ (\ref{14}) and apply
$\zeta(\beta,\eta)$ multiplicatively
to the shower distribution $\cal S$, but not to the soft distribution \tt,
since the effect of
degradation is already included in the determination of \cc\ through
the use of the experimental
values of $dN_{\rm ch}/d\eta(\beta)$ in Eq.\ (\ref{8}). Equation
(\ref{11a}) is used for $T(\beta, \eta)$.
For $\cal S$$\cal S$\ recombination
we do not apply
$\zeta(\beta,\eta)$ quadratically, since the shower partons are from one jet.

With the degradation effect taken into account the results on the
ratio $R_{CP}$ are shown in Fig.\
7. We note that there is improvement in the agreement with data compared to
Fig.\ 6, especially at $\eta=2.2$. At $\eta=1.0$ the solid  line no
longer overshoots the dashed
line at high \pt. There is no change at $\eta=0$ since
$\zeta(\beta,0)=1$, and there is
an improvement of the fit at $\eta=3.2$.  Since the data are for
either $(h^++h^-)/2$ or $h^-$, while our calculation is for $\pi^+$
specifically, perfect agreement between theory and experiment should
not be expected. The effect of momentum
degradation is at most 30\% on the \pt\ distributions at $\eta=2.2$,
so such changes are barely
perceptible in the log plots of the spectra in Fig.\ 2, which
therefore remain as the prediction of
our treatment. $R_{CP}$ in linear scale reveals the degradation
effect more sensitively.
It is evident from Fig.\ 7 that the essence of forward production of
pions in d+Au collisions is
essentially captured in our description of hadronization by parton
recombination, when $T$ is allowed to depend on \bb\ and \et, and
when the effects of momentum degradation is taken into account.

\section{Asymmetry Ratio for Backward to Forward Rapidities}

So far we have restricted our study to only the forward region. The
backward region on the gold side
contains the properties of the nucleus not present at large \et, and
should therefore behave
differently from what we have obtained in previous sections. Recent
data from STAR show significant
asymmetry in the ratio of the charged hadron spectra for backward to
forward rapidities in the range
$0.5<|\eta|<1.0$ \cite{ja2}. That ratio reaches a peak higher than
1.3 for $2<p_T<3$ \gev. At a
qualitative level the phenomenon can easily be understood in our
approach to the problem, since there are more soft partons at
$\eta<0$ than at $\eta>0$. We now want to examine the asymmetry
quantitatively as another test of our treatment of hadronization.

The formalism for particle production at $\eta<0$ is the same as for
$\eta>0$. Both \tt\ and $\cal S$\
must, however, change, as \et\ enters the negative region. The
parametrizations of the hard parton
distributions $f_i(k_T,y)$ are already given in Table II. Since the
data for the backward region
are for the range $-1.0<\eta<-0.5$, we shall use the parameters for
$y=-0.75$ in Table II.
Similarly, the values for $y=0.75$ will be used for the forward
region. For the soft parton
distribution \tt$(p_1)$ we continue to use Eq.\ (\ref{8}) to
determine \cc\ with $dN_{\rm
ch}/d\eta(\beta)$ taken directly from the data \cite{rn}. For \bb\ =
0-20\%, we obtain
$C(\beta,-0.75)=12.372$ and $C(\beta,0.75)=11.527$ (GeV/c)$^{-1}$.
This represents a small, but
significant, asymmetry of the soft parton density. As for the \et\
dependence of the inverse slope $T(\beta, \eta)$ we use the formula,
Eq.\ (\ref{11a}), already determined in Sec.\ 3 from the region
$\eta>0$, now applied to $\eta=\pm 0.75$.

We now can calculate the
\pt\ distributions for $\pi^+$ using the appropriate values of $C, T$
and $f_i$ at $\eta=\pm 0.75$.  The results are, of
course, not visually distinguishable from that at $\eta=0$ in Fig.\ 2
plotted in log scale. However,
   their ratio plotted in linear scale is more sensitive to the small
changes. The backward/forward asymmetry ratio is defined by

\begin{eqnarray}
R_{B/F}(p_T, |\eta|)={dN_{\pi}/dp_T\,d\eta\,(\eta=-|\eta|)\over
dN_{\pi}/dp_T\,d\eta\,(\eta=+|\eta|)}\ .  \label{16}
\end{eqnarray}
The data for $R_{B/F}(p_T, 0.75)$ as shown in Fig.\ 8
are for all charged hadrons \cite{ja2}. Thus our calculation for
$\pi^+$ alone is not enough for the purpose of comparison with the
currently available data. In the same way that we have treated proton
production in Au+Au \cite{hy} and d+Au collisions \cite{hy4}, we
calculate the proton spectrum at $\eta=\pm0.75$. Furthermore, we take
the $\pi^-$ yield to be the same as for $\pi^+$, and $\bar p$ yield
to be 0.7 of that of $p$. The sum of $\pi^++\pi^-+p+\bar p$ is shown
by the solid line in Fig.\ 8. The result has the correct rise for
$p_T<2$ GeV/c, but is lower than the data at higher \pt. We expect
that the production of kaons can further increase the theoretical
curve, but since they involve strange quarks that are enhanced in the
soft component, we do not digress here to that peripheral subject.
Our present result from the non-strange sector is sufficient to
indicate that the asymmetry data can be understood in our approach to
hadronization. Note that our result on $R_{B/F}$ has been obtained
without any new free parameter. A better way to compare theory with
data would be to have identified pions at $\eta=\pm0.75$, which is
within the feasibility of some RHIC experiment.

The conclusion that one can draw from this study of the
backward-forward asymmetry is that there is
no transition of basic physics from multiple scattering in ISI on the
$\eta<0$ side to gluon
saturation on the $\eta>0$ side \cite{lm}. Our emphasis on the
hadronization process in the final
state provides a universal framework for the description of particle
production at all \et, \bb, and
\pt.

Based on saturation physics, a recent calculation of the
low-\pt\ distribution of charged hadrons in minimum bias d+Au
collisions at $\eta=3.2$ appears to have good agreement with data,
when only the scattering of $q\bar q$ dipole on the nucleus described
as CGC is taken into account \cite{jjm}. However, the contribution of
gluons to the cross section is not negligible, since the
density of gluons in a proton at $x=0.25$ is more than half the
density of the quarks at $Q\le5$ \gev\ \cite{zebu}.  Furthermore, hadronization by use of fragmentation function
is subject to the usual question about the $p/\pi$ ratio that has
shown the inadequacy of the fragmentation model.  Thus the agreement
with data at low \pt\ seems fortuitous.

It has recently been stated
that the Cronin effect seen at midrapidity goes away at forward
rapidities because the supposed origin of the enhancement at $\eta=0$
(i.e., multiple scattering in the initial state) is replaced by CGC
that is responsible for the suppression at $\eta>0$ \cite{lm,jb,jjm}.
The results of this work combined with those of Refs.\ \cite{hy2,hy4}
suggest that neither mechanisms are the primary causes of the
enhancement and suppression, and that both phenomena are the consequences of the same hadronization process by recombination.

\section{Conclusion}

On the basis of parton recombination we have successfully described
pion production at intermediate
\pt\ in d+Au collisions in the forward rapidity region. The formalism
is an extension of the one at
midrapidity where the Cronin effect has been explained in terms of
FSI only, and where the
experimental fact $R_{CP}^p>R_{CP}^{\pi}$ is interpreted as a
consequence of the dominance of 3-quark recombination over
fragmentation. In the extension to $\eta>0$ only one new parameter,
$\varepsilon$, is introduced to describe the \bb\ and \et\ dependence
of $T$, but
no new physics has been added. The suppression of
$R_{CP}$ at $\eta>1$ is due mainly
to the reduction of the density of soft partons that recombine either
among themselves or with the
semi-hard shower partons. The effect of momentum degradation
responsible for baryon stopping has
been considered, and is found to have a minor effect on $R_{CP}$, although it does
 render a better agreement with  the data at all $\eta$ and \pt. The reduction of soft
parton density in the forward direction may be related to gluon
saturation, but there is no explicit reliance on small-$x$ physics in
the calculation. At large \et\
the production of hard partons is suppressed, so pQCD calculations
supplemented by fragmentation is
likely to underestimate the pion spectra at intermediate \pt.

Extending the consideration to the backward region, we have used the
same dependence  of the inverse slope $T(\beta, \eta)$ on \et, now
extrapolated from $\eta>0$ to $\eta<0$. We find that  the general
properties of the backward-to-forward ratio of the charged hadron \pt\
distributions can be reproduced by our calculation of only the
non-strange sector. Thus, particle production at all \et\ can
be described by the same formalism for all \pt\ at any centrality.
There is no need for any change
of physics in going from the backward to the forward region.

To further verify the validity of our treatment, the predicted pion
spectra in Fig.\ 2 should be checked by experiments.  To have the correct $R_{CP}$, as in Fig.\ 7, is only the necessary condition for the underlying physics to be relevant, but not sufficient.  Spectra themselves may disagree with the data, yet still have their ratio  come out right.
Proton spectra should also be measured at
$\eta>0$ and the $p/\pi$ ratio
shown as a function of \pt, \bb, and \et. Although we have not
calculated the \pt\ distribution of proton for $\eta>1$, we expect that the $p/\pi$ ratio would not be small for $\eta>1$, although lower than the maximum of the
ratio for $\eta=0$ because the lower soft
parton density at $\eta>1$ inhibits the formation of protons more
than it does pions.  The measurement of that ratio will provide a severe
test on any model of particle production at any \et.

Another area of investigation that can shed light on forward physics
is to determine the presence or
absence of back-to-back jets in azimuthal correlation when a particle
is detected at large \et\ and
intermediate \pt. If the \pt\ distributions of the various components
contributing to the pion
spectrum at $\eta=3.2$ in Fig.\ 3 is correct, then we do not expect
any significant jet signature
until $p_T>3$ \gev.  Even at $p_T=4$ \gev\ the dominant component is
\tt\tt\ recombination, so at
$\Delta\phi=\pi$ there should be only a small jet-like component that
stands above a high level of
background uniform in $\phi$. Some aspect of that feature has already
been observed in the
preliminary data of STAR \cite{ao}.

Since hadronization of partons by recombination is a process in the final stage of evolution of the partons, it is not in conflict with any dynamical model that correctly describes the beginning and subsequent evolution of those partons. Thus our work here provides the necessary link between the predicted parton spectra and the observed hadronic data for testing the validity of any proposed model.

\section*{Acknowledgment}
We are grateful to  E.\ Gonzalez, F.\ Matathias, M.\ Murray, R.\
Nouicer, A.\ K.\ Purwar, L.\ Ruan, R.\ Venugopalan, G.\ Veres, N.\ Xu, and Z.\ Xu for
helpful discussions, communications and advanced sharing of data.  This work
was supported, in part,  by the U.\ S.\ Department of Energy under Grant No. DE-FG02-96ER40972 and DE-FG02-87ER40328,  and by the Ministry of Education of China under Grant No. 03113, and National Natural Science Foundation of China under Grant No. 10475032.

\newpage

\centerline{\large{\bf Figure Captions}}
\vskip0.5cm
\begin{description}

\item Fig.\ 1. Transverse momentum distributions of $\pi^+$ produced
at midrapidity in dAu collisions for three centrality cuts. Light solid lines are for
the recombination from the \tt\tt\ component, dashed lines for
\tt$\cal S$\ and dashed-dotted lines for $\cal S$$\cal S$\ components. Heavy solid
lines are the sums of all three components.

   \item Fig.\ 2. Transverse momentum distributions of $\pi^+$ produced
at different pseudorapidities for two centrality cuts.

\item Fig.\ 3. Transverse momentum distributions of $\pi^+$ produced
at \et\ = 3.2 for 0-20\% centrality, showing the three components.

\item Fig.\ 4. Distribution of gluons produced in central d+Au
collisions at four pseudorapidities.

\item Fig.\ 5. $R_{CP}$ for 0-20\%/60-80\% (filled circles and solid
lines) and 30-50\%/60-80\% (open circles and dashed lines) for four
pseudorapidities when $T$ is assumed to be constant. Data are from
\cite{ia}. No momentum degradation is considered in the calculation.

\item Fig.\ 6. Same as for Fig.\ 5 but with $T(\beta, \eta)$ given
by Eqs.\ (\ref{11a}) and (\ref{12a}).

\item Fig.\ 7. Same as for Fig.\ 6 but with momentum degradation considered.

\item Fig.\ 8. Ratio of \pt\ distributions for backward to forward
pseudorapidities at $|\eta|=0.75$. Data are for all charged hadrons
from \cite{ja2}. Solid line is the calculated ratio for
$\pi^++\pi^-+p+\bar p$.

\end{description}

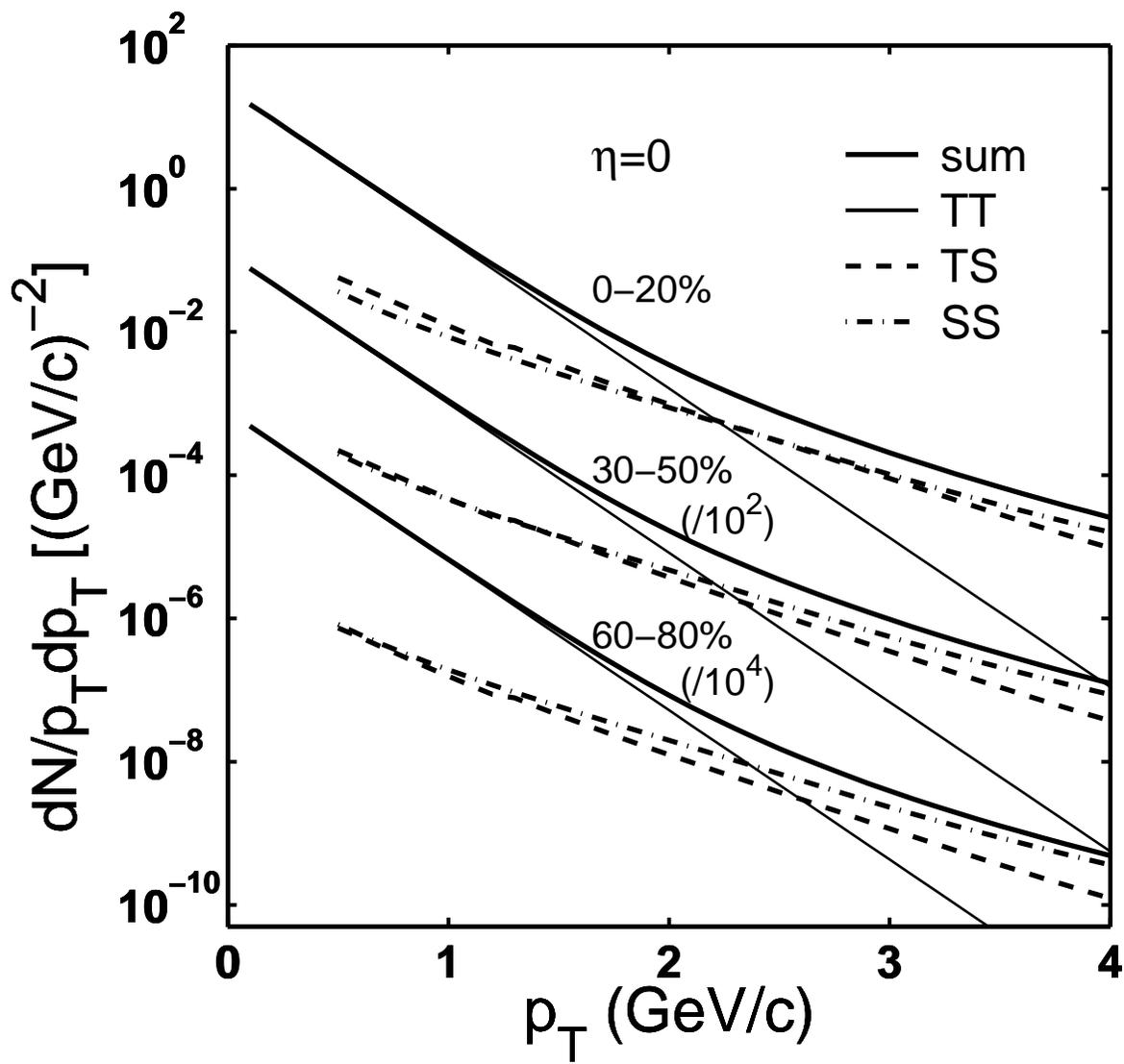
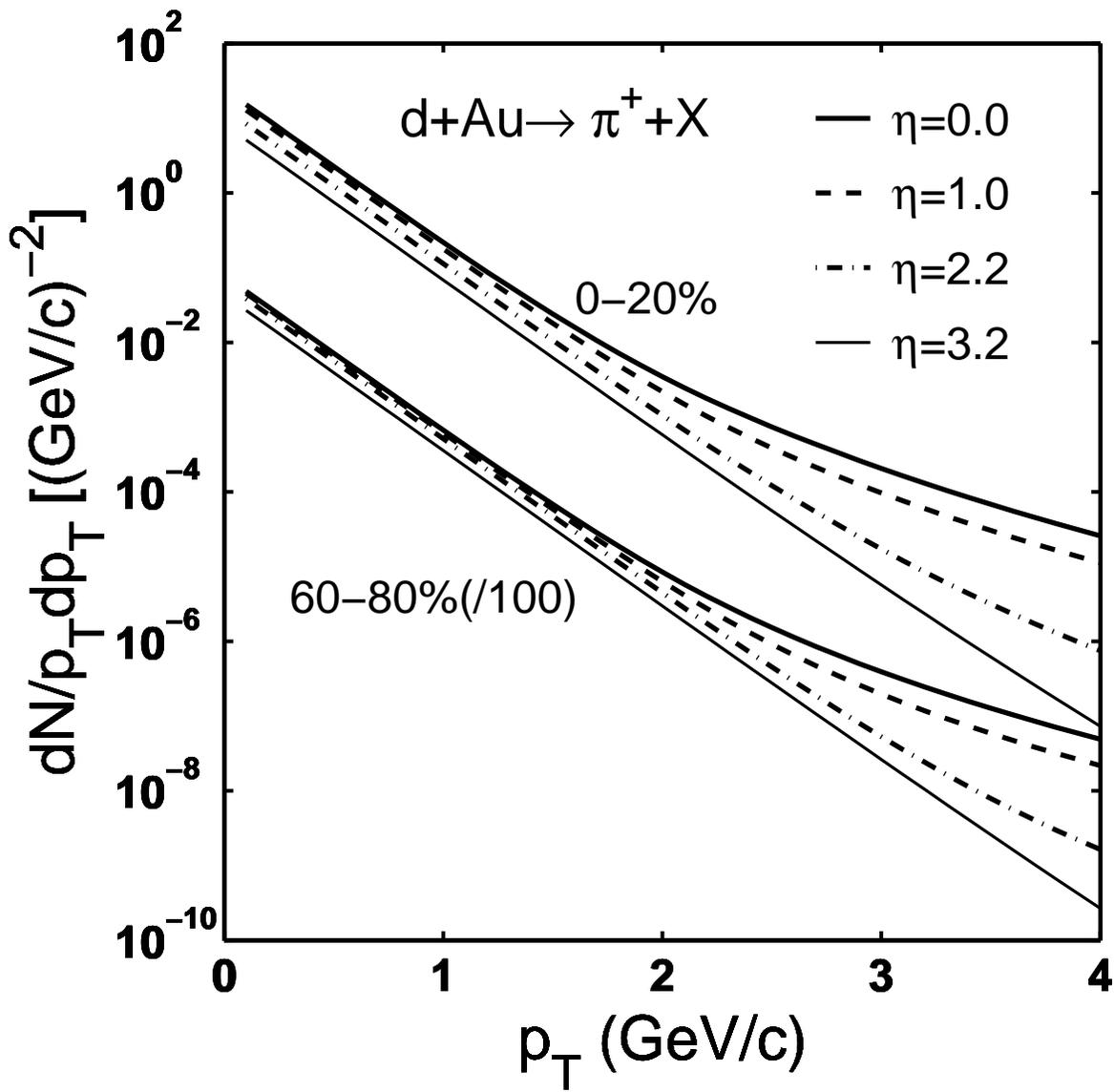
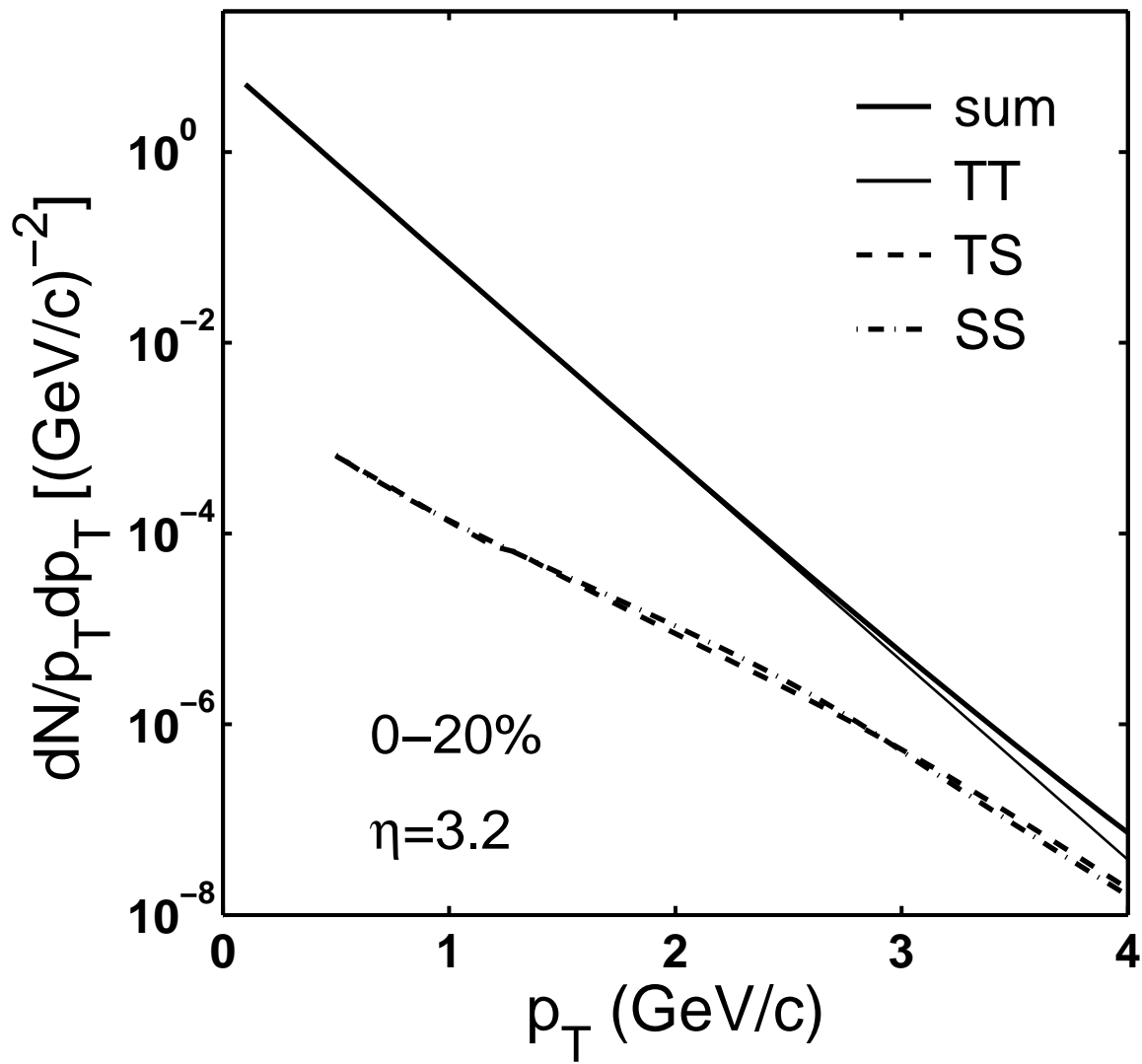
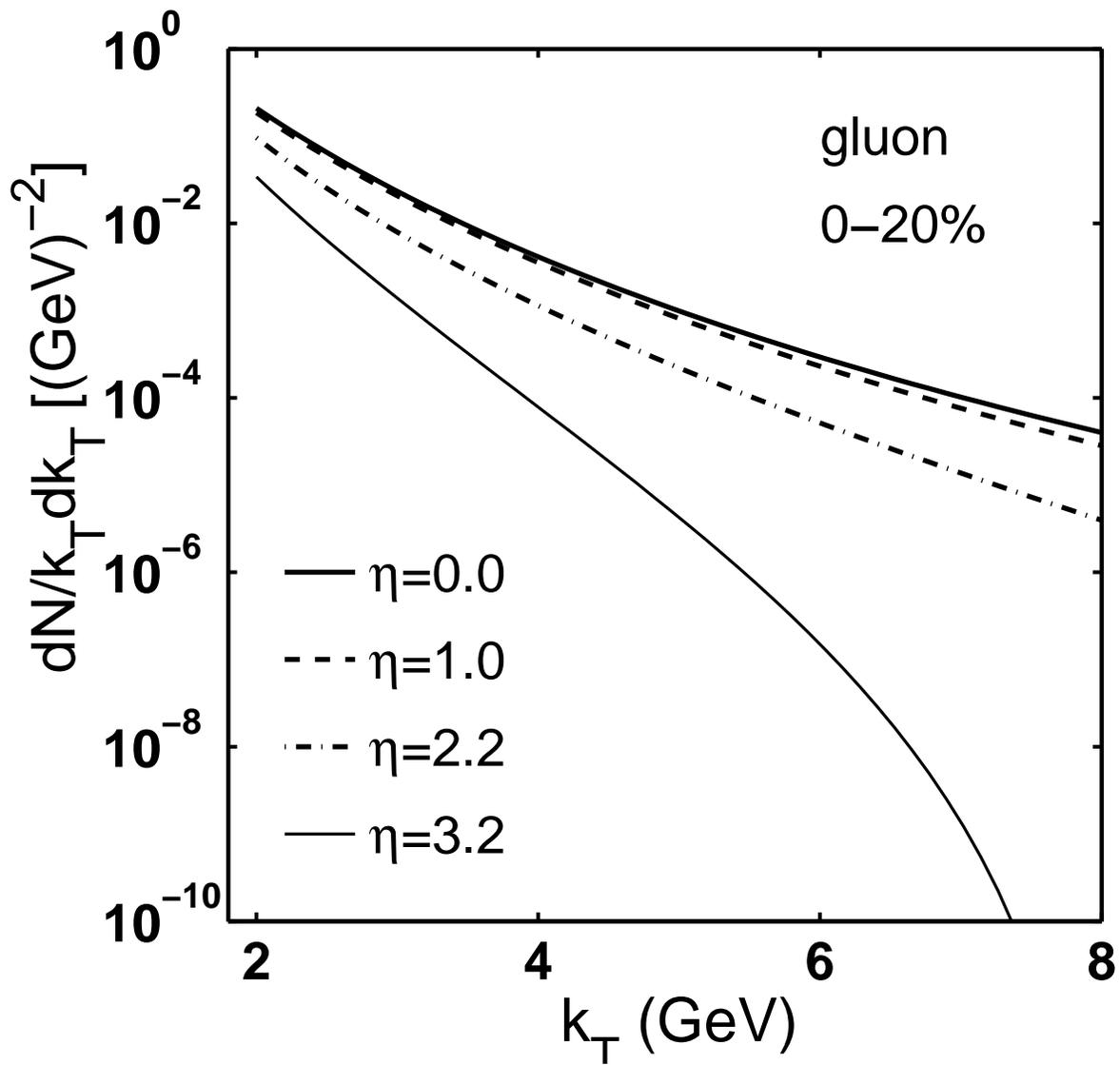
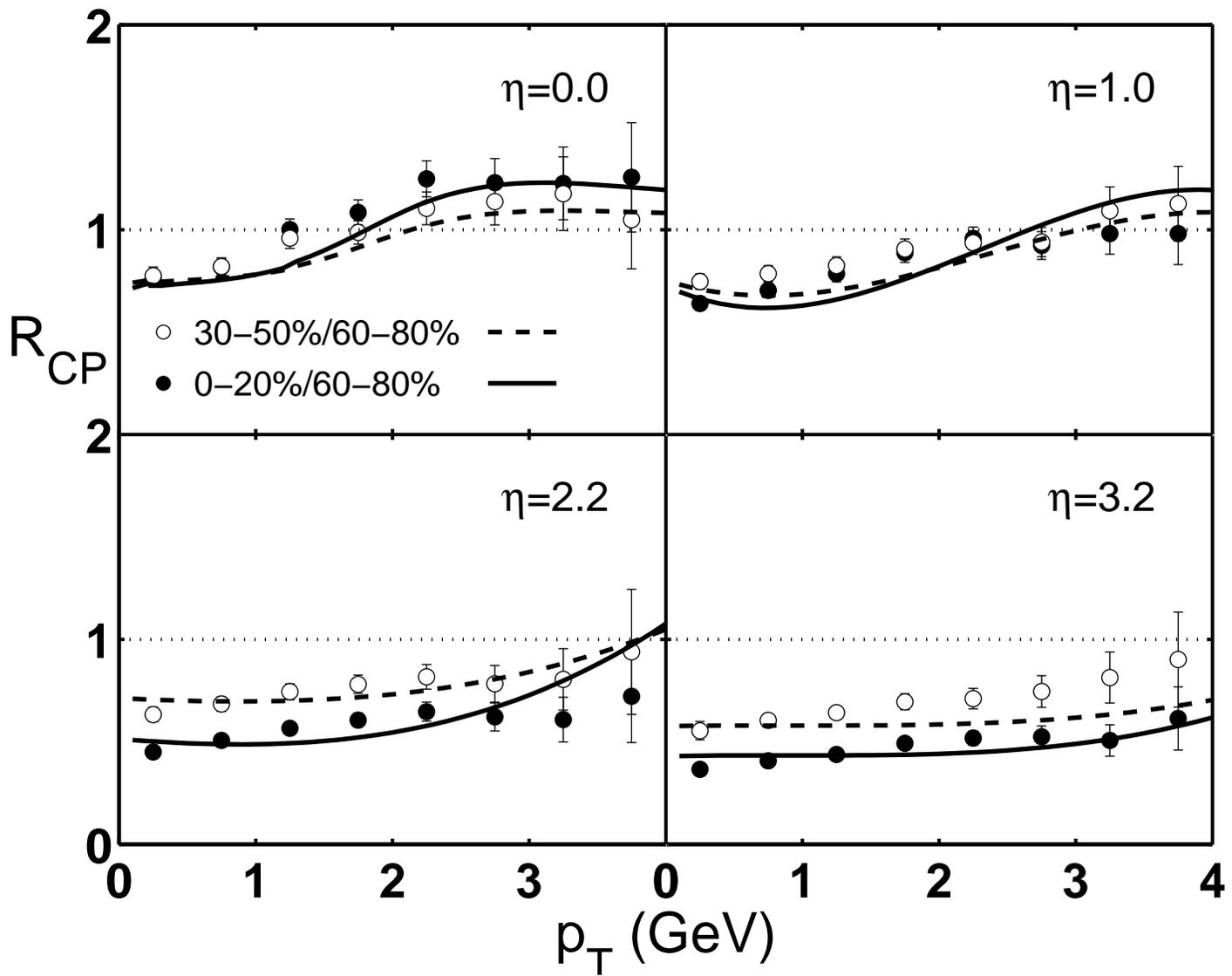
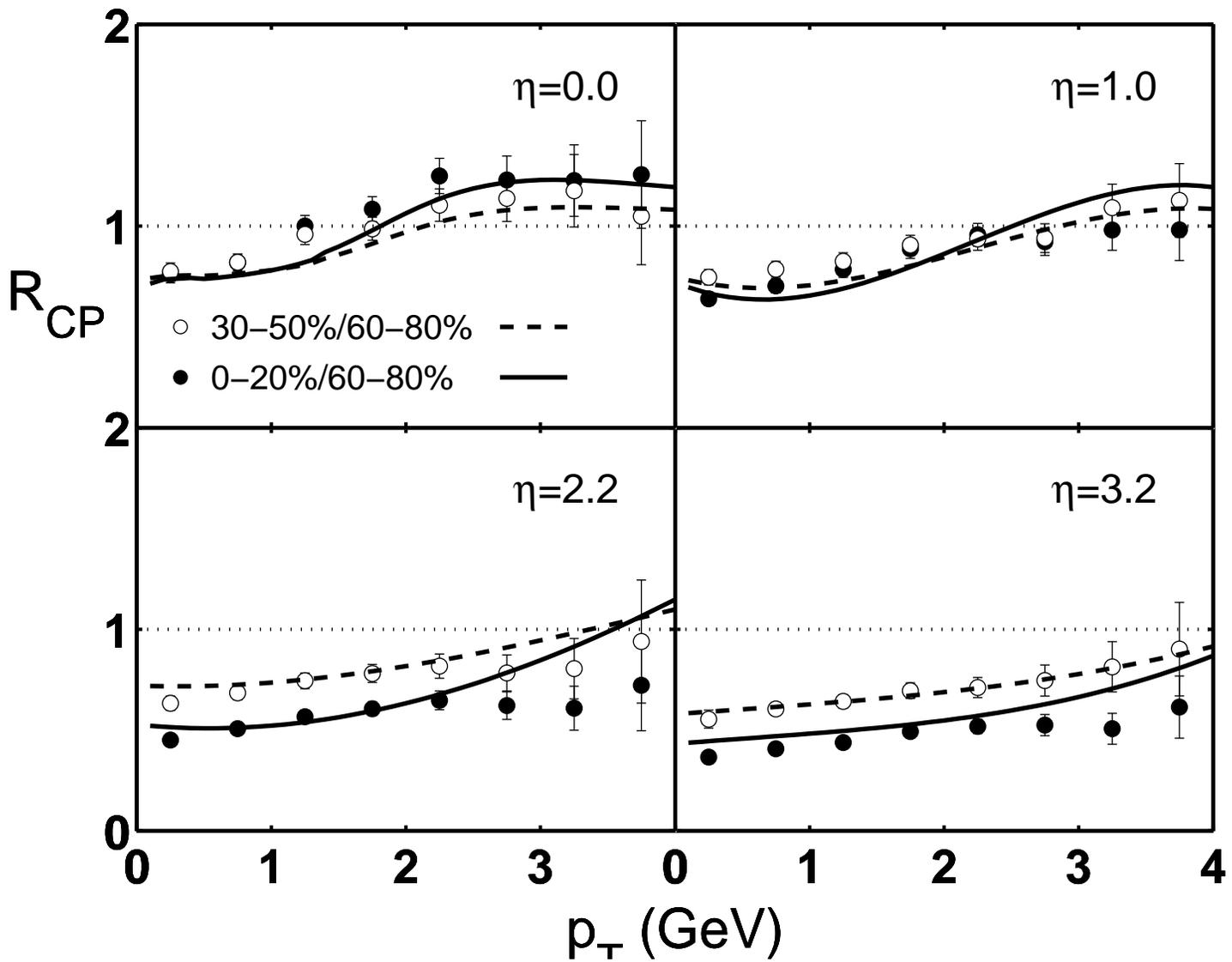
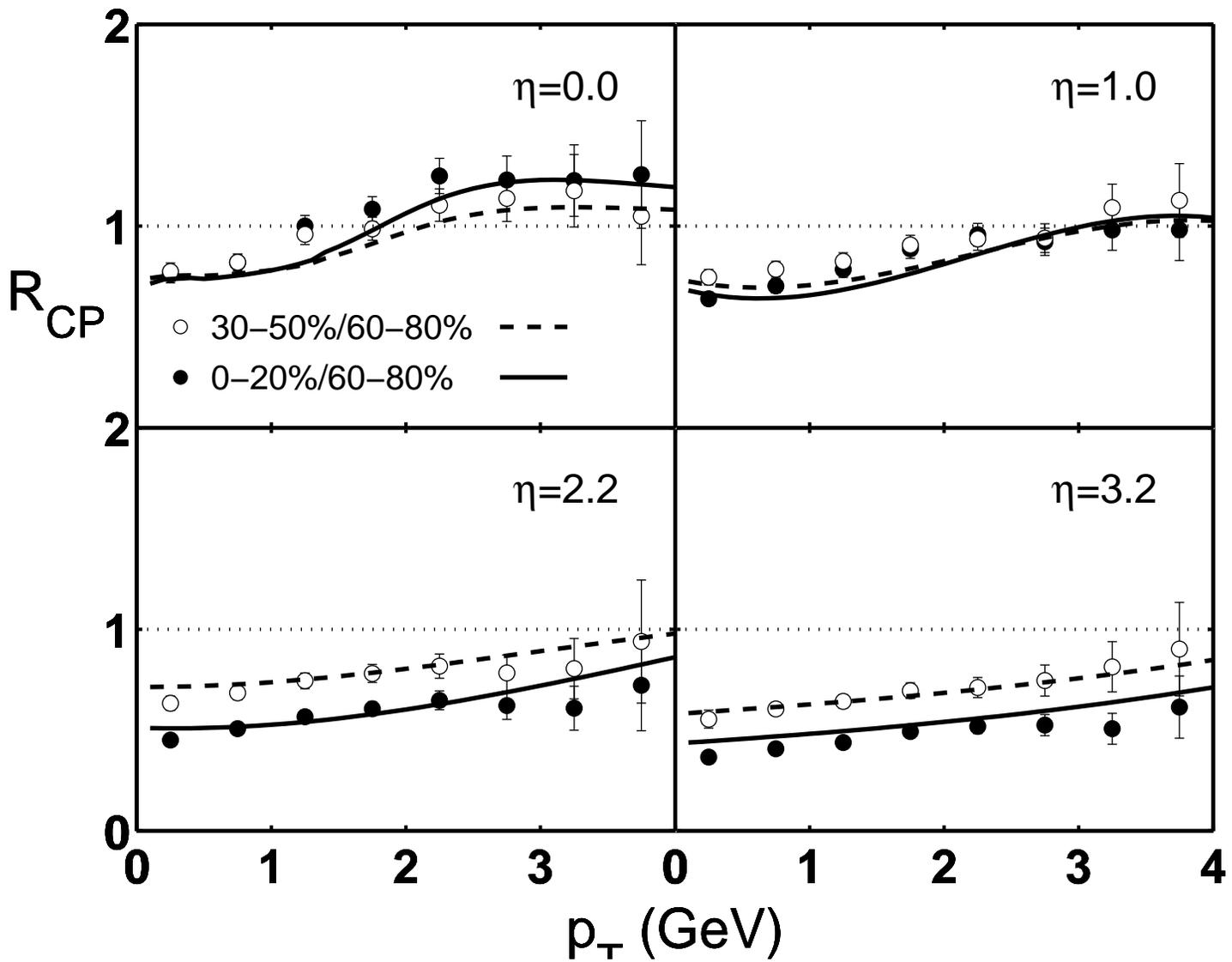
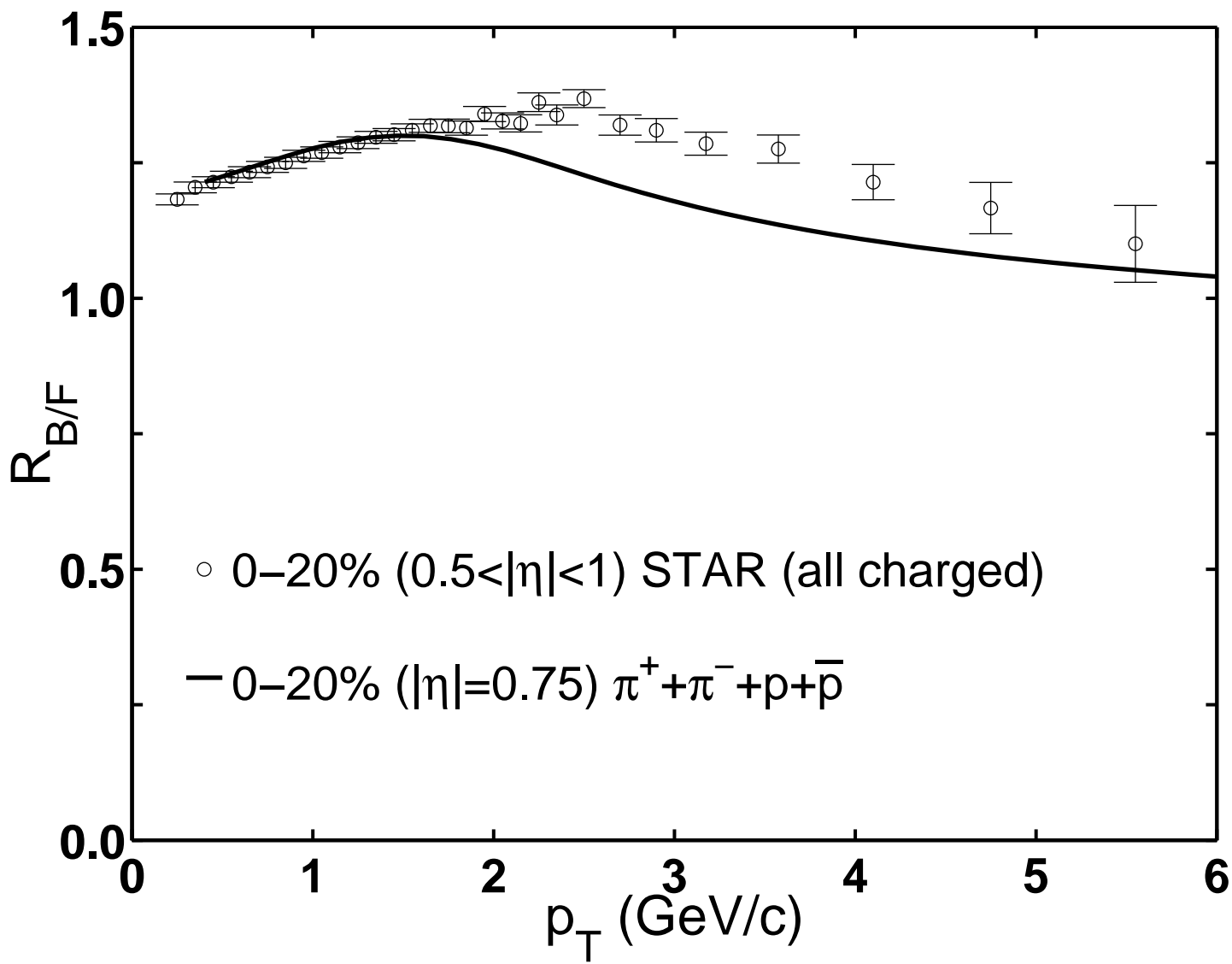
\end{document}